\documentclass{ctr}

\usepackage{ctrfont}
\usepackage{natbib}

\usepackage{url}
\usepackage{amsmath}
\usepackage{amssymb}
\usepackage{graphicx}
\usepackage{psfrag}
\usepackage{epsfig}
\usepackage{appendix}

\title{Parallel variable-density particle-laden turbulence simulation}
\shorttitle{Variable-density particle-laden turbulence}
\author{Hadi Pouransari, Milad Mortazavi \and Ali Mani}
\shortauthor{Pouransari et al.}
\begin{document}
\maketitle
%========================================================================
%=============================INTRODUCTION==============================
%========================================================================
\section{Motivation and objectives}
\label{sec:intro}

Variable density flows are ubiquitous in a variety of natural and industrial systems. Multi-phase flows in natural and industrial processes, astrophysical flows, and flows involved in combustion processes are such examples. For an ideal gas subject to low-Mach approximation, variations in temperature can lead to a non-uniform density field. In this work, we consider radiatively heated particle-laden turbulent flows. We have developed a parallel C++/MPI-based simulation code for variable-density particle-laden turbulent flows (which can be downloaded from \url{https://hadip@bitbucket.org/hadip/soleilmpi}). The code is modular, abstracted, and can be easily extended or modified.

We were able to reproduce well-known phenomena such as particle preferential concentration \citep{eaton1994preferential}, preferential sweeping \citep{wang1993settling}, turbulence modification due to momentum two-way coupling \citep{squires1990particle, elghobashi1993two}, as well as canonical problems such as the Taylor-Green vortices and the decay of homogeneous isotropic turbulence.

This code is also able to simulate variable-density flows. Therefore, we were able to study the effect of radiation on the system. We reproduced one of our previous results, using a different code assuming constant density and Boussinesq approximation, on radiation-induced turbulence \citep{zamansky2014radiation}. We also reproduced the results of the homogeneous buoyancy-driven turbulence \citep{batchelor1992homogeneous}, where the Boussinesq approximation is relaxed. Moreover, using the Lagrangian-Eulerian capabilities of the code, many new physics have been studied by different researchers. These include spectral analysis of the energy transfer between different phases \citep{pouransari2014spectral}, subgrid-scale modeling of particle-laden flows \citep{urzaycharacteristic}, settling of heated inertial particles in turbulent flows \citep{frankel2014settling}, and comparisons between Lagrangian and Eulerian methods for particle simulation subject to radiation \citep{vie2015particle}.

This code is developed to simulate a particle-laden turbulent flow subject to radiation. Each of these three elements can be removed from the calculation. In the simplest case, the code can be used as a homogeneous isotropic turbulence (HIT) simulator. We implemented two different geometrical configurations as explained in section \ref{sec:bound}.

Particles are simulated using a point-particle Lagrangian framework. Particles are two-way coupled with fluid momentum and energy equations using linear interpolation and projection. The fluid is represented through a uniform Eulerian staggered grid. Spatial discretization is second-order accurate, and time integration has a fourth-order accuracy. The fluid is assumed to be optically transparent. However, particles can be heated with an external radiation source. Hot particles can conductively heat their surrounding fluid. Therefore, the fluid density field is variable. This leads to a variable coefficient Poisson equation for hydrodynamic pressure of the flow. We have developed a novel parallel linear solver for the variable density Poisson equation that arises in the calculation. Governing equations and numerical methods are explained in sections \ref{sec:model} and \ref{sec:numerics}, respectively.
%==========================================================================
%=============================PHYSICAL MODEL==============================
%==========================================================================
\section{Physical model}
\label{sec:model}
In this section we provide a set of equations for the background flow and particles that are solved in the code. The primary variables for the flow are density ($\rho$), momentum ($\rho {u_f}_i$), and temperature ($T_f$), and those for particles are location ($x_p$), velocity (${u_p}_i$), and temperature ($T_p$).

Mass and momentum conservation equations for gas are as follows
%%%%%%% CONSERVATION OF MASS %%%%%%%%%
\begin{equation} \label{eqn:COM}
\frac{\partial \rho}{\partial t}+\frac{\partial}{\partial x_j}(\rho {u_f}_j)=0,
\end{equation}
%%%%%% CONSERVATION OF MOMENTUM %%%%%%%%
\begin{multline} \label{eqn:NS}
\frac{\partial} {\partial t}(\rho {u_f}_i)+\frac{\partial}{\partial x_j}(\rho {u_f}_i {u_f}_j)=-\frac{\partial p}{\partial x_i}+\mu\frac{\partial}{\partial x_j}\left(\frac{\partial {u_f}_i}{\partial x_j} +\frac{\partial {u_f}_j}{\partial x_i} -\frac{2}{3}\frac{\partial {u_f}_k}{\partial x_k}\delta_{ij}\right) \\
+ A \rho {u_f}_i + g_i (\rho - \rho_0) + {\cal{P}}\left( \frac{{u_p}_i- {\cal{I}}({u_f}_i) } { \tau_p } \right).
\end{multline}
The last term in Eq. (\ref{eqn:NS}) represents the momentum transfer from the particles to the fluid.
The operator $\mathcal{P}$ projects data from particle location to the fluid grid(per unit volume). $\mathcal{I}$ represents interpolation from the Eulerian grid to a particle location. In the current version of the code, both $\mathcal{P}$ and $\mathcal{I}$ are implemented using a linear approximation.
The term $A \rho {u_f}_i$ corresponds to the linear forcing \citep{rosales2005linear}. $g_i$ in $g_i (\rho - \rho_0)$ is the gravitational acceleration in the $i^{\mbox{th}}$ direction after subtracting the static pressure. Last three terms in Eq. (\ref{eqn:NS}) can be switched on/off for different calculations. $\tau_p$ is the particle inertial relaxation time.

Conservation of energy for the gas phase is described by the following equation
%%%%%%% CONSERVATION OF ENERGY %%%%%%%%
\begin{equation} \label{eqn:GEE}
\frac{\partial}{\partial t} (\rho C_v T_f)+\frac{\partial}{\partial x_j}(\rho C_p T_f {u_f}_j)=k \frac{\partial^2 T_f}{\partial x_j \partial x_j}+{\cal{P}}\left(2\pi D_p k \left(T_p- {\cal{I}}(T_f)\right)\right).
\end{equation}
The last term in Eq. (\ref{eqn:GEE}) represents the heat transfer from particles to the fluid. $D_p$ is the particle diameter, and $k$ is the gas thermal conductivity coefficient.

The thermodynamics of the system is governed by the ideal gas equation of state $\rho R T_f =P_0$, where $P_0$ is the thermodynamic pressure.
Spatial pressure variation in the equation of state is neglected. This is justified by the low-Mach assumption.

A point-particle Lagrangian framework is used for particles; therefore, particle location is governed by the following equation
%%%%%%%% PARTICLE KINEMATICS %%%%%%%
\begin{equation} \label{eqn:KIN}
\frac{d {x_p}_i}{dt} = {u_p}_i.
\end{equation}
Spherical particles with a small particle Reynolds number based on the slip velocity in a dilute mixture are considered.
Assuming a large density ratio $\rho_p/\rho_f\gg1$ and particle diameter smaller than the Kolmogorov length scale, the Stokes drag is the most important force on the particles \citep{maxey1983equation}. Hence, the dynamics of the particles is governed by the following equation
%%%%%% PARTICLE DYNAMICS %%%%%%%%
\begin{equation} \label{eqn:DYN}
\frac{d {u_p}_i}{dt}=-\frac{{u_p}_i- {\cal{I}}({u_f}_i) } { \tau_p }.
\end{equation}

The following equation describes the conservation of energy for each particle;
%%%%%%% PARTICLE ENERGY EQN %%%%%%%%
\begin{equation} \label{eqn:PEE}
\frac{d}{dt} (m_p {C_v}_p T_p) = \frac{1}{4}\pi D_p^2 I \epsilon_p - 2\pi D_p k (T_p - {\cal{I}}(T_f)).
\end{equation}
In the above equation, $m_p = \rho_p \pi D_p^3 /6$ and ${C_v}_p$ are, respectively, particle mass and heat capacity. The first term in the right-hand side of Eq. (\ref{eqn:PEE}) is the heat absorption by a particle with emissivity $\epsilon_p$ in an optically thin medium with uniform radiation intensity $I$. The second term represents heat transfer to the fluid.
%====================================================================
%=============================NUMERICS==============================
%====================================================================
\section{Numerics} \label{sec:numerics}
\subsection{Spatial discretization} \label{sec:disc}
\begin{figure}
	\centering
  \includegraphics[width=0.6\textwidth]{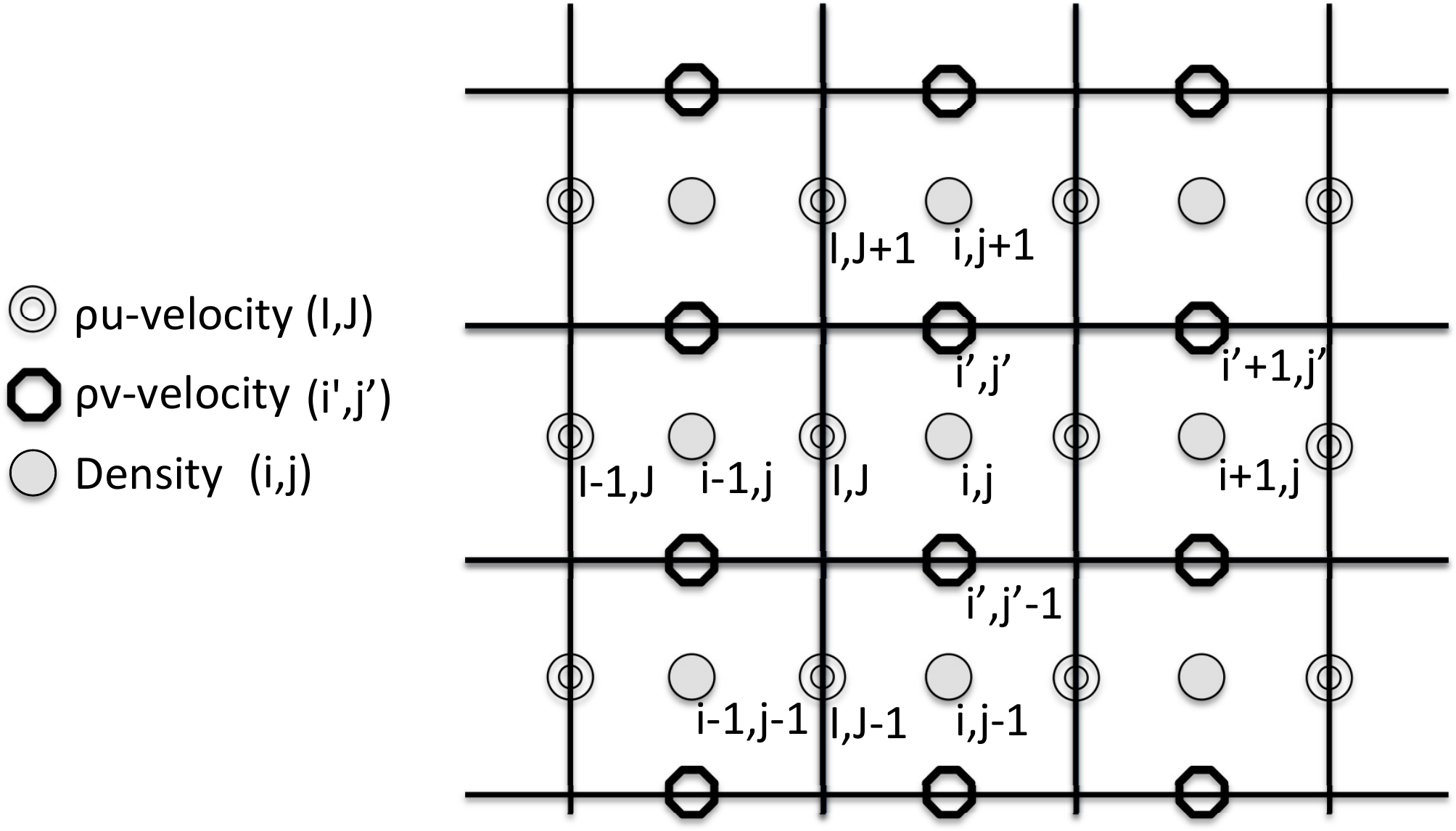}
  \caption{An illustration of a 2D uniform staggered grid, with momentum stored on cell faces, and scalar quantities at cell centers.}
\label{fig:SCHEMATIC}
\end{figure}

We have used a uniform three-dimensional staggered grid for the fluid in this code. A sketch of a two-dimensional uniform staggered grid is shown in Figure \ref{fig:SCHEMATIC}.
Our discretization preserves the total mass, momentum, and kinetic energy. Therefore, for an inviscid flow, we enforce to conserve the total kinetic energy, as shown in Figure \ref{fig:cons}. The details of the energy conservative numerical method used in the code are explained in appendix \ref{sec:appB}.

\begin{figure}
\centerline{\includegraphics[width = 0.9 \textwidth]{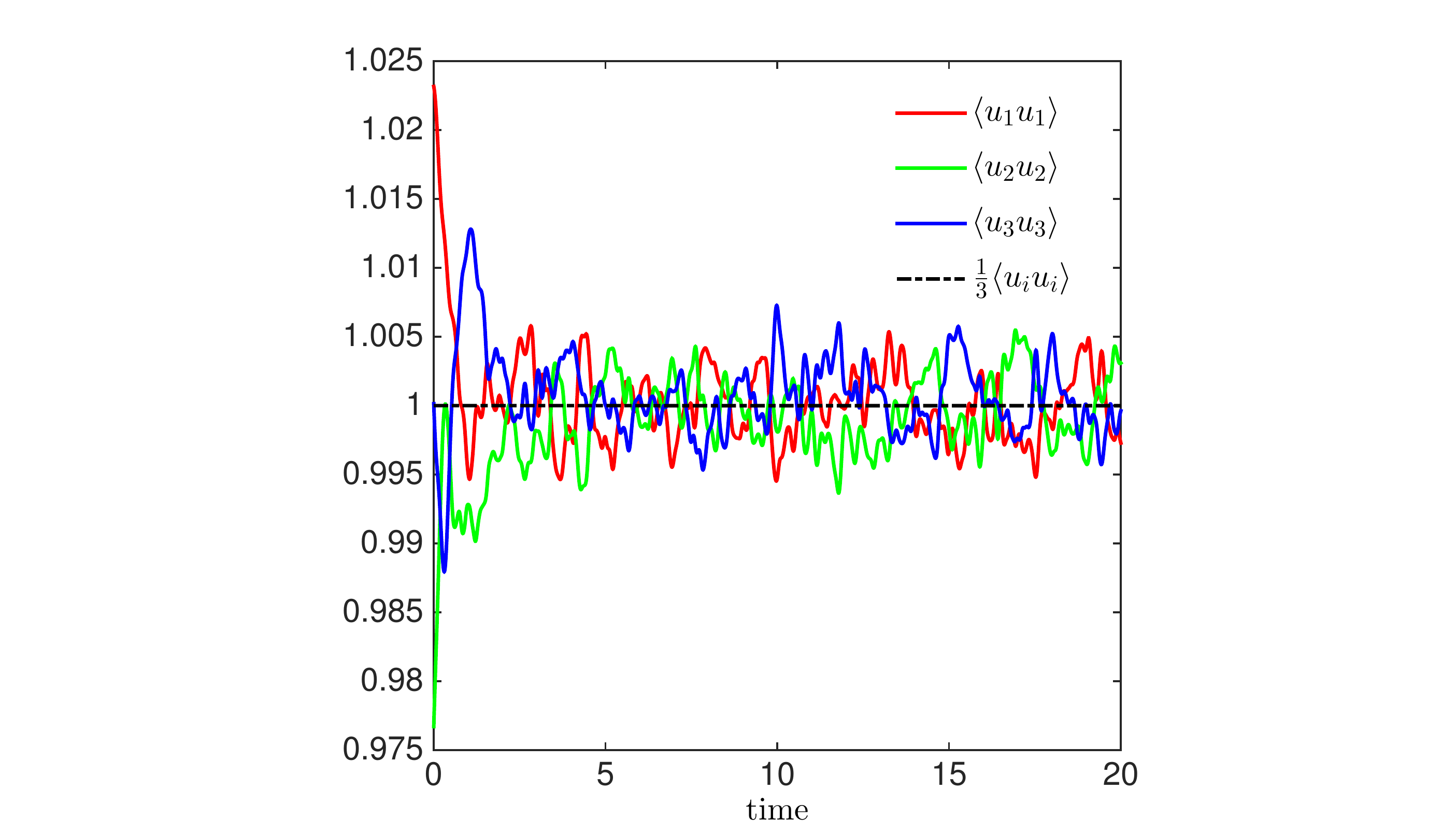}}
\caption{\label{fig:cons}Conservation of the total kinetic energy for an inviscid calculation. Note that energy is being transferred from different components of the velocity, while the total kinetic energy is preserved.}
\end{figure}

\subsection{Boundary and initial conditions} \label{sec:bound}
This code supports two different configurations: 1) Triply periodic, which is a box with periodic boundary conditions (BC) in all directions. 2) Inflow-outflow, which is a duct with periodic BC in the $y-z$ plane, and convective BC in the $x$ direction
\begin{equation}
\label{eqn:convBC}
\frac{\partial}{\partial t} (.)+U_x \frac{\partial}{\partial x} (.)=0.
\end{equation}

\begin{figure}
\centerline{\includegraphics[width = 0.9 \textwidth]{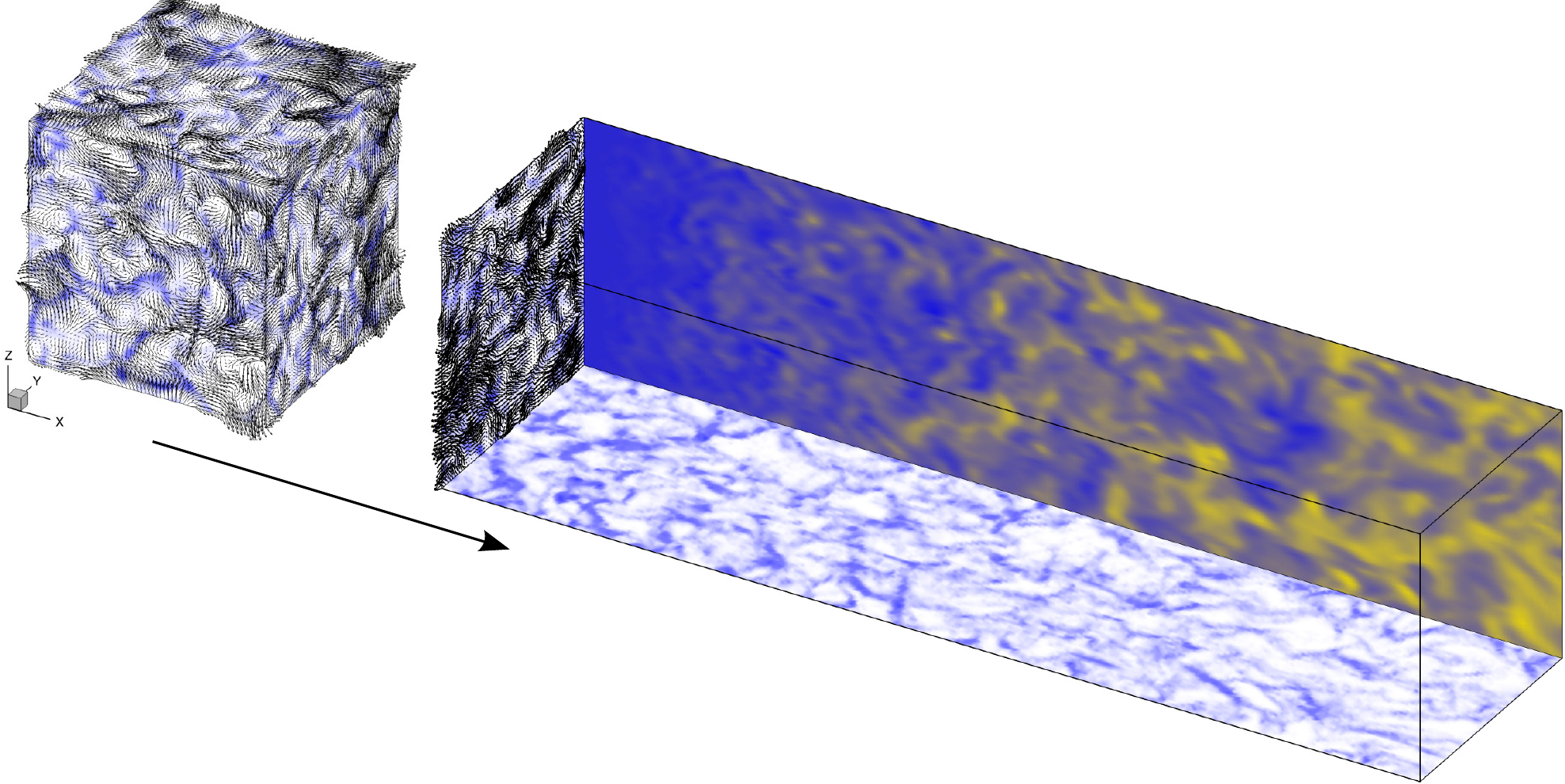}}
\caption{\label{fig:DOMAIN}Inflow-outflow computational domain. Left: the triply periodic box used as a turbulence generator for the inflow-outflow domain (right). Particle concentration is shown on the sides of the box, and the bottom side of the inflow-outflow domain. Gas density contours are shown on the rear side of the inflow-outflow domain.}
\end{figure}

The inflow comes from an HIT simulation, sustained with linear forcing, which is running simultaneously with the duct simulation. At each time-step, the last slice of the HIT simulation is being copied to the first slice of the inflow-outflow simulation. The computational domain is shown in Figure \ref{fig:DOMAIN}.

The code accepts various initial conditions for particles and flow. In particular, we have provided a script to generate initial turbulence using a Passot-Pouquet spectrum \citep{passot1987numerical}.

\subsection{Time advancement} \label{sec:time}
The time integration is performed using a fourth-order Runge Kutta (RK4) scheme for both particles and the flow. We first solve for momentum without considering the pressure gradient term. Using the divergence implied by the energy equation (note that the flow filed is not divergence free when radiation is on), we solve the variable coefficient Poisson equation and correct the fluid momentum field. The energy equation involves an energy transfer term between particles and fluid. Therefore, in order to accurately calculate the divergence condition for the fluid pressure, we need to have energy transfer from particles at the next substep before solving the Poisson equation. In other words, at each step the right-hand side of the particle energy equation is calculated at the next substep. However, the right-hand sides of all other equations are evaluated at the current substep. Details of the numerical scheme are explained in appendix \ref{sec:appA}.

\subsection{Poisson equation} \label{sec:Poisson}
Solving for a variable density flow requires solving a variable coefficient Poisson equation for the hydrodynamic pressure as follows
\begin{equation}
\nabla. \left( \frac{1}{\rho} \nabla p \right)= f,
\end{equation}
where $f$ is determined using the energy equation. Consider $p = p_g + \delta p$, where $p_g$ is our estimated value from the previous step. Then the above equation can be re-written as
\begin{equation} \label{eqn:Pois}
\nabla^2 \delta p = \rho f - \rho \left( \nabla \frac{1}{\rho} \right) \cdot \left( \nabla p_g \right) - \nabla^2 p_g - \rho \left( \nabla \frac{1}{\rho} \right) \cdot \left( \nabla \delta p \right).
\end{equation}
All terms in the right-hand side of the above equation are known except for the last one. We ignore the last term and iterate to update $p_g \leftarrow p_g + \delta p$, until convergence. The convergence rate depends on the variation in the density field. Eq. (\ref{eqn:Pois}) represents a constant coefficient Poisson equation (after ignoring the last term) that is solved using a parallel Fourier transform in periodic directions \citep{plimpton1997particle} and a parallel tri-diagonal solver in the non-periodic direction.
%=====================================================================
%=============================SOFTWARE==============================
%=====================================================================
\section{Software description} \label{sec:soft}
\subsection{Software Architecture}
In this section we briefly describe the significance and abilities of each class:

\verb!params!: This class reads all input parameters from a file. After some initial calculations (before the simulation begins), all physical parameters can be accessed through this class.

\verb!tensor0!: Stores a local 3D grid (i.e., for one process). Simple arithmetic, spatial discretization, interpolation, etc., are implemented in this class. Similarly, the class \verb!tensor1!  stores a local grid of 3D vectors. In other words, \verb!tensor1! encompasses three instances of \verb!tensor0!.

\verb!gridsize!: stores and provides the logical information (e.g., size in each direction) for a local grid.

\verb!proc!: has logical information (e.g., rank, neighbors, etc.) for each process.

\verb!communicator!: takes care of updating halo cells for the grid by communicating between processors. All other parallel algorithms are implemented within this class.

\verb!poisson!: solves the variable coefficient Poisson equation.

\verb!particle!: holds information of $N_p$ particles. Also, all particle-related algorithms are implemented in this class.

\verb!grid!: represents the full grid, including all scalar and vector quantities for the flow.

Time integration loops are implemented in \verb!simulation.cpp!. Note that with RK4, every quantity $Q$ has four instances: $Q$ (value at time-step $n$), $Q_{int}$ (value at substep $k$), $Q_{new}$ (value at substep $k+1$), and $Q_{np1}$ (value at time-step $n+1$).

\subsection{Software Functionalities}
\label{sec:func}
Using different classes defined in the code, one can easily compute various calculus operations (differentiations, interpolations, etc.) in parallel, without working directly with any MPI-related command. Even though all parallel implementations are under the hood, they can be modified within the \verb!communicator! class. The code is parallelized in three directions. The number of processes in each direction is determined in the \verb!proc! class.

Flow and particle statistics can be computed at a desired frequency and stored as binary files (using parallel I/O). Numerical techniques are implemented completely modular; therefore, modifying the code to employ new numerics needs minimal effort (e.g., going from second-order to a fourth-order stencil).

After starting each simulation, a summary of information is stored in a file named \verb!info.txt!. Simulation can be stopped, and all data be saved before the designated time by changing a flag in the file \verb!touch.check!.
%=======================================================================
%=============================VERIFICATION==============================
%=======================================================================
\section{Verifications}
\subsection{Decay of Taylor-Green vortices}
\begin{figure}\centering
\centering
\includegraphics[width = 0.8 \textwidth] {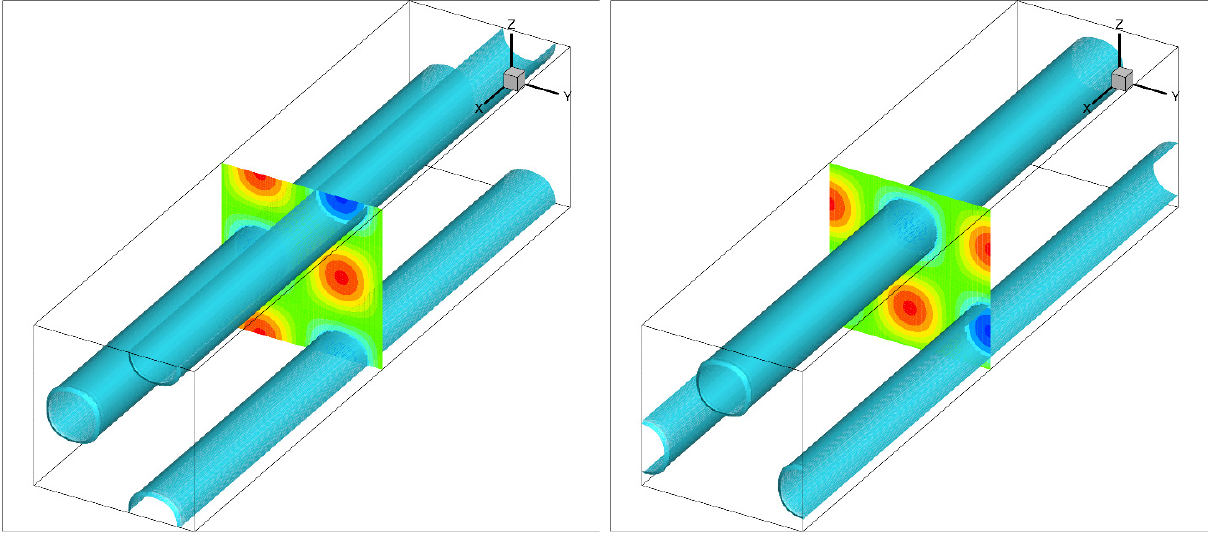}
\caption{\label{fig:PRScatter}$v$ contours and iso-surface (left); $w$ contours and iso surface (right). }
\end{figure}

\begin{figure}\centering
\centering
\includegraphics[width = 0.9 \textwidth]{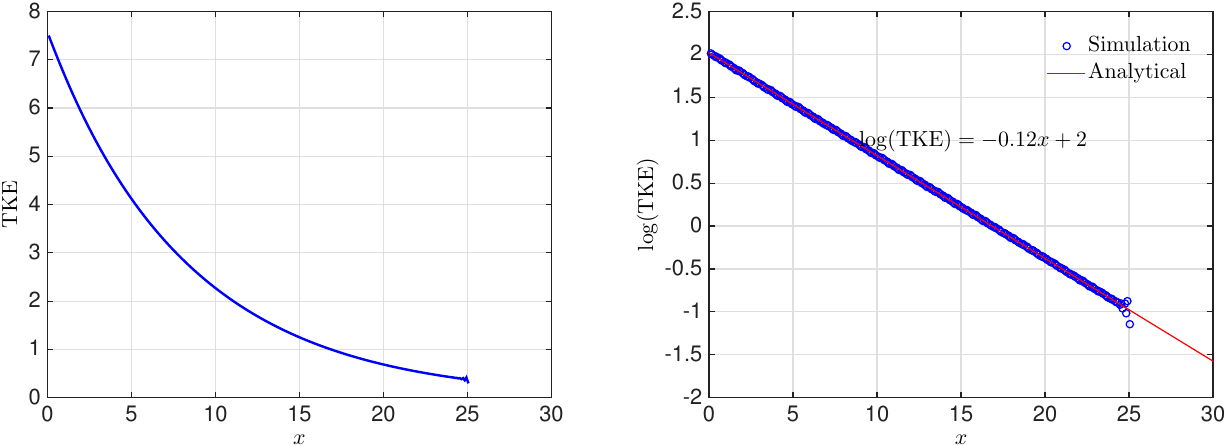}
\caption{Decay of the TKE in x-direction (left). Linear fitting on the semi-log plane (right).}
\label{fig:TGdecay}
\end{figure}

\begin{figure}\centering
\centering
\includegraphics[width = 0.45 \textwidth]{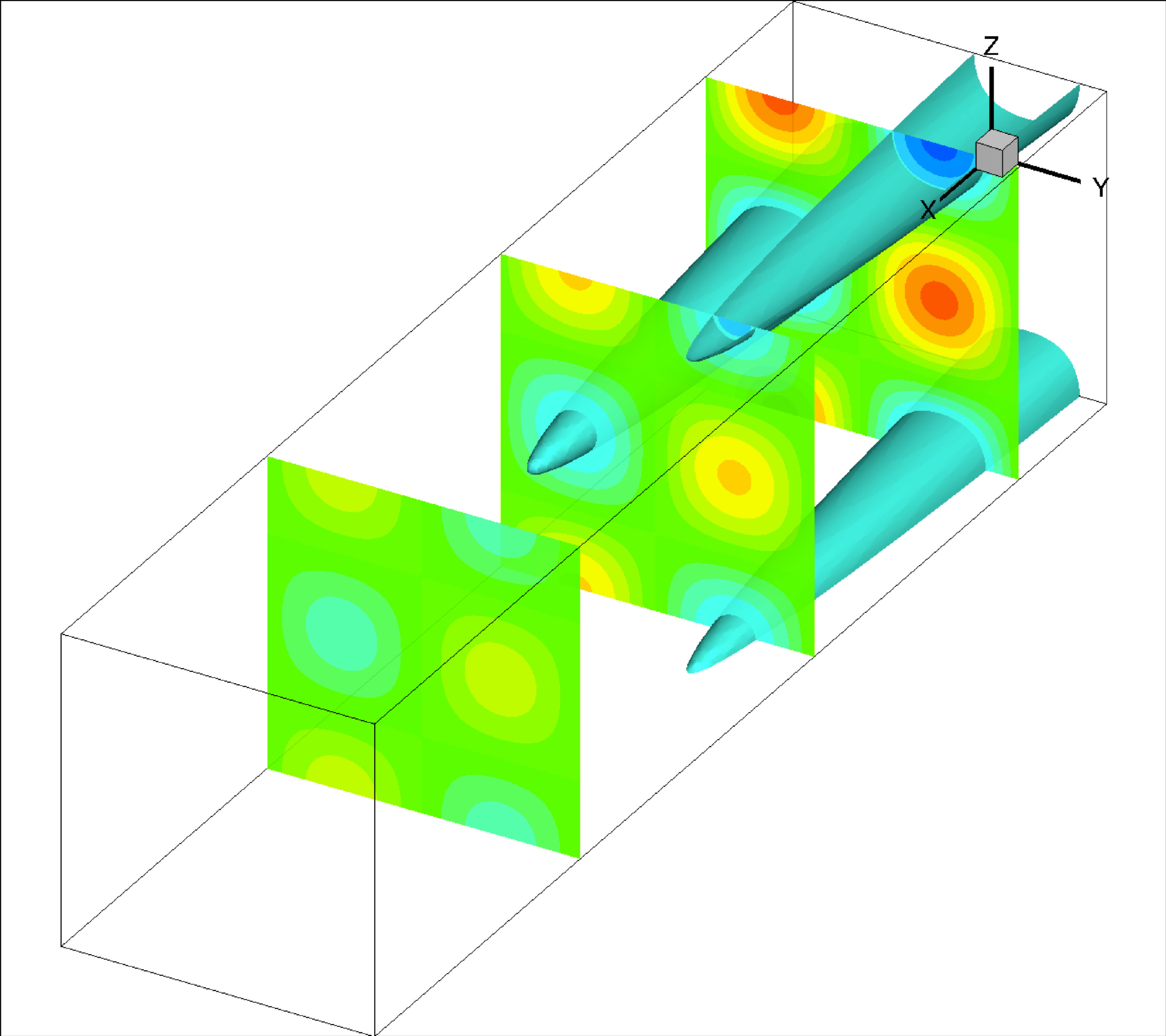}
\caption{$v$ contours and iso-surface}
\label{fig:Viso}
\end{figure}

We have verified the carrier-flow implementation using the Taylor-Green vortices, an analytical solution to the Navier-Stokes equations. Having set the initial condition of the triply periodic box simulation as
\begin{equation}
\begin{split}
u(x,y,z)&=U_x,\\
v(x,y,z)&=\sin(y) \cos(z),\\
w(x,y,z)&=-\cos(y) \sin(z),
\end{split}
\end{equation}
the velocity field will maintain its two-dimensional shape as shown in Figure \ref{fig:PRScatter}.
\\
We can also look at the decay rate of the Taylor-Green vortices in time. Starting with the initial velocity field given by the above equations, we expect an exponential decay rate for the turbulent kinetic energy $\sim e^{(2A-4\nu)t}$. Note that $A$ is the linear forcing coefficient as in Eq. \ref{eqn:NS}, which is only applied to the box simulation. A slice that enters the duct at time $t_0$ has turbulent kinetic energy (TKE) $\sim e^{(2A-4\nu)t_0}$. However, since there is no linear forcing in the duct, the decay rate of this slice scales like $e^{-4\nu t}$. In order to follow this decay we should consider the convective velocity in $x$ direction, $U_x$ (which is $U_x=10$ in this case). It means that at time $t_0+\Delta t$ the slice location is moved by $U_x \Delta t$. In other words, a slice located at position $x$ has decayed from its enterance value for $\Delta t=x/U_x$ time units. Thus, by looking at one snapshot at an arbitrary time, $t$, we expect the following distribution for the TKE
$$
\mbox{TKE} \sim e^{-4\nu \frac{x}{U_x}} \cdot e^{(2A-4\nu)t_0} = e^{-4\nu \frac{x}{U_x}} \cdot e^{(2A-4\nu)(t-\frac{x}{U_x})}.
$$
In the above expression, $e^{-4\nu \frac{x}{U_x}}$ is the decay that occurs in the duct, and $e^{(2A-4\nu)(t-\frac{x}{U_x})}$ is the TKE at the time at which this slice is copied from the box.
Hence, $\mbox{TKE} \sim e^{\frac{2A}{U_x}x} = e^{.12x}$ (here, A=.6 is used). The decay of TKE is shown in Figure \ref{fig:TGdecay}, and the $v$-component of the velocity iso-surface is illustrated in Figure \ref{fig:Viso}. At the very end of the domain, due to a first-order accurate convective BC, there is a source of error, as is clear from Figure \ref{fig:TGdecay}.


\subsection{Buoyancy driven turbulence}
\begin{figure}\centering
\centering
\includegraphics[width = 0.45 \textwidth]{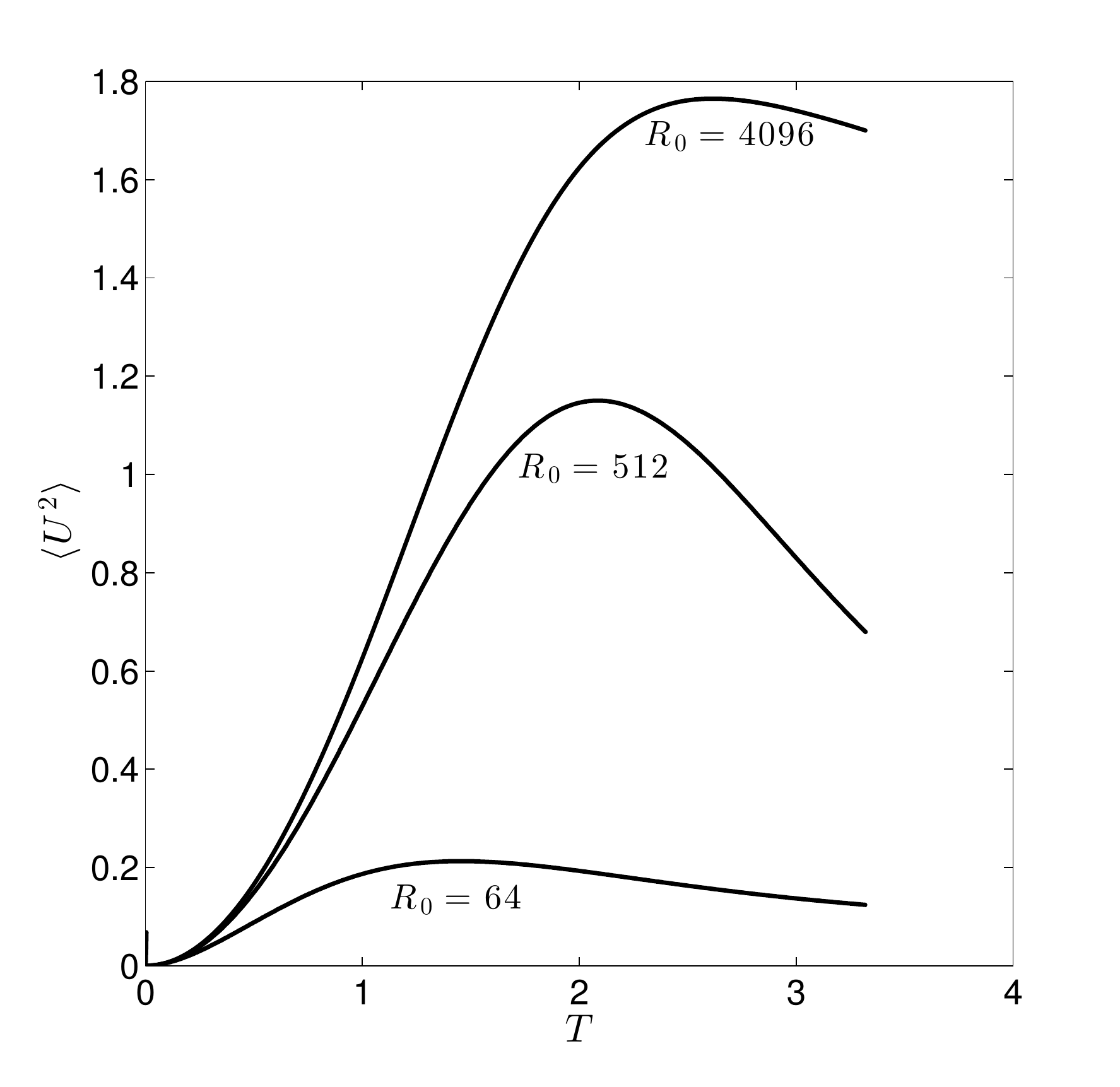}
\caption{Generation of homogeneous buoyancy-driven turbulence. Non-dimensionalization is similar to that of the original paper by \cite{batchelor1992homogeneous}.}
\label{fig:TKEBatch}
\end{figure}

Another sample problem, against which we tested the code is the homogeneous buoyancy-generated turbulence flow \citep{batchelor1992homogeneous}. In the original paper, the Boussinesq approximation is used, which is less accurate than a variable density calculation. The flow starts at rest, while the initial density field is non-homogeneous, and is drawn from an artificial spectrum. When gravity is present, the inhomogeneity of the density field results in the generation of turbulence. The turbulence decays eventually. These results are shown in Figure \ref{fig:TKEBatch}. The results are matched with those presented by \citep{batchelor1992homogeneous} for smaller Reynolds numbers. For large Reynolds numbers (corresponding to larger variations in the gas density field) the results are slightly different. This can be explained by inaccuracy of the Boussinesq approximation when density variation is relatively large.

\subsection{Capturing particle preferential concentration}
\begin{figure}\centering
\includegraphics[width=0.65\textwidth]{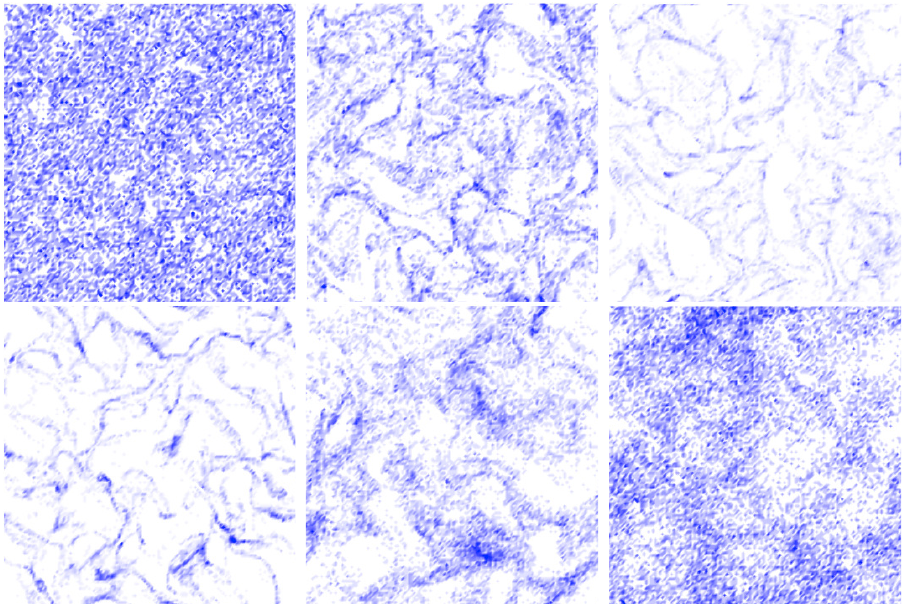}
\caption{2D slices of the three-dimensional particle-laden turbulence simulation for Kolmogorov-based Stokes number = $0.0625,~0.25,~0.5,~1,~4$, and $8$, respectively, from top left to bottom right. The slice thickness is $1/128$ of the length of the box.}
\label{fig:snapshot_me} 
\end{figure}

In a particle-laden turbulence, the spatial distribution of the dispersed phase deviates from the homogeneous distribution \citep{eaton1994preferential}. This is due to the interaction of particles with turbulent eddies.
Particle Stokes number (ratio of the particle relaxation time to the turbulent Kolmogorov timescale) is the relevant non-dimensional number that determines the inhomogeneity of particles.
The highest preferential concentration is expected to appear for the Stokes number of order unity. In Figure \ref{fig:snapshot_me}, snapshots of particle distribution for different Stokes numbers are shown, confirming maximum preferential concentration at Stokes number 1. A quantitative analysis of particle preferential concentration is provided in \citep{pouransari2015}.

\subsection{Effect of domain partitioning}
\begin{figure}\centering
\centering
\includegraphics[width = 0.5 \textwidth]{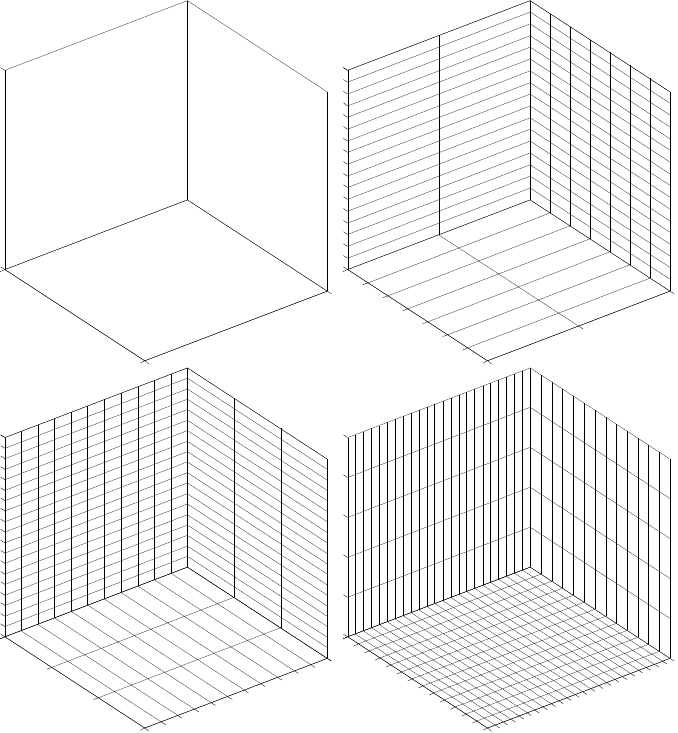}
\caption{Different domain decompositions for parallelization.}
\label{fig:decompos}
\end{figure}

We used various domain decompositions (for parallelization), and with the same initial condition, measured the difference between solution (momentum, density, particle location/velocity, etc.) after 10 time-steps. We found that solutions are equal up to machine precision. Some of the domain decompositions that we used are shown in Figure \ref{fig:decompos}.
%======================================================================
%=============================CONCLUSION==============================
%======================================================================
\section{Conclusions}
\label{sec:conclusion}
In this paper, we introduced a variable-density particle-laden turbulent flow simulation code. The code is fourth-order accurate in time, and second-order accurate in space. It is fully parallel using MPI. Particles are modeled as Lagrangian points, while fluid is represented using a uniform staggered Eulerian grid. This code has demonstrated the expected results for various canonical problems, and has been used to discover physics in a variety of new fields involving particles, flow, and radiation.
%=============================================================================
%=============================ACKNOWLEDGMENTS==============================
%=============================================================================
\section*{Acknowledgments}
This work was supported by the United States Department of Energy under the Predictive Science Academic Alliance Program 2 (PSAAP2) at Stanford University.
%=====================================================================
%=============================APPENDIX A==============================
%=====================================================================
\appendix
\section{} \label{sec:appA}
In this section, we explain the details of the numerics used to solve for a variable-density flow with low-Mach number assumption.

Consider $C_{RK4}^{post} = [\frac{1}{6},\frac{1}{3},\frac{1}{3},\frac{1}{6}]$ and $C_{RK4}^{pre} = [\frac{1}{2},\frac{1}{2},1]$ as constant coefficients used to implement the RK4 time integration. Also, assume $k=0,1,2,3$ corresponds to the RK4 four substeps. Assume $Q$ is some quantity of interest (e.g., density). At each RK4 substep $Q_{new}$ is obtained from $Q^{(n)}$ and $Q_{int}$. $Q^{(n)}$ is the value of $Q$ at time-step $n$, $Q_{int}$ is the value at the previous RK4 substep, and $Q_{new}$ is the new value of $Q$ after the RK4 substep. For the first RK4 substep we initiate $Q_{int}=Q^{(n)}$.
Note that we are solving for conservative fields $\rho$ and $\rho u_i$ (here, we dropped the fluid subscript for simplicity). In the following, the time integration steps are explained.

We start by solving for the flow density
\begin{equation}
\rho_{new}=\rho^{(n)}-C_{RK4}^{pre}[k] \cdot \Delta t \cdot  \frac{\partial}{\partial x_j} (\rho u_j)_{int}.
\end{equation}
Next, solve for momentum without pressure
\begin{equation}  \label{eqn:mom}
(\rho u_i)_{new} = \widetilde{(\rho u_i)}_{new} - C_{RK4}^{pre}[k] \cdot \Delta t \cdot \left( \frac{\partial p}{\partial x_i} \right)_{int},
\end{equation}
\begin{equation}
\widetilde{(\rho u_i)}_{new}=(\rho u_i)^{(n)}+C_{RK4}^{pre}[k] \cdot \Delta t \cdot \widetilde{\text{RHS}}_{int}^{mom},
\end{equation}
where
\begin{multline}
\widetilde{\text{RHS}}^{mom}=\frac{\partial}{\partial x_j}\left(\mu(\frac{\partial u_i}{\partial x_j}+\frac{\partial u_j}{\partial x_i}-\frac{2}{3}\frac{\partial u_k}{\partial x_k}\delta_{ij})\right)+A \rho u_i\\+(\rho-\rho_0) g_i - \frac{\partial}{\partial x_j}(\rho u_i u_j) + {\cal{P}}\left( \frac{{u_p}_i- {\cal{I}}({u_f}_i) } { \tau_p } \right).
\end{multline}
Here, $\widetilde{(.)}$ refers to the value of some quantity without considering the pressure gradient.

We then compute divergence of the velocity at the next substep\\
\begin{equation} \label{eqn:divUnew}
\nabla . \vec{u}_{new} = \frac{k }{C_p} \nabla^2 (\frac{1}{\rho_{new}}) + \frac{\alpha R}{C_p} \frac{1}{{P_0}_{new}} - \frac{C_v}{C_p} \frac{1}{{P_0}_{new}} \left( \frac{d P_0}{dt} \right)_{new},
\end{equation} 
where $\alpha = {\cal{P}}\left(2\pi D_p k (T_p- {\cal{I}}(T_f))\right)$ is the energy transferred from the particles to the gas per unit time per unit volume. Note that in the above equation ${P_0}_{new}$ and $\left( \frac{d P_0}{dt} \right)_{new}$ are not known. The last term, $\frac{C_v}{C_p} \frac{1}{{P_0}_{new}} \left( \frac{d P_0}{dt} \right)_{new}=\chi$, is a scalar (only function of time) and will be determined later to satisfy the consistency of the Poisson equation. In order to compute ${P_0}_{new}$, we can obtain the time-evolution equation for $P_0$, which is assumed to be only a function of space with a low-Mach number assumption. To do this, one can integrate the energy equation in space, after which all divergence terms vanish due to periodicity. Hence,
\begin{equation} \label{eqn:dP0dt}
\frac{d P_0}{dt}=\frac{\alpha R \bar{c}}{C_v}.
\end{equation}
Note that for the inflow-outflow case $P_0$ is constant in time and space.
The right-hand side of the above equation is a constant number (in time and space). The time evolution-equation for $P_0$ using RK4 is
\begin{equation}
{P_0}_{new}={P_0}^{n}+C_{RK4}^{pre}[k] \cdot \Delta t \cdot \frac{\langle \alpha \rangle R}{C_v}.
\end{equation}
As explained earlier, we keep $\chi$ as an unknown to be determined later in order to make the Poisson equation well-posed. \footnote{Theoretically, the values resulting from Eq. (\ref{eqn:dP0dt}) should be consistent with the Poisson equation. However, we make this adjustment in order to overcome the ill-posedness resulting from the numerical errors.}
Rewrite Eq. (\ref{eqn:divUnew}) as below:
\begin {equation}
\label{eqn:divNewGen}
\nabla . \vec{u}_{new} = \text{RHS}^{div}_{new}- \chi,
\end {equation}
where in the above equation:
$$
\text{RHS}^{div}_{new}= \frac{k }{C_p} \nabla^2 (\frac{1}{\rho_{new}}) + \frac{\alpha R}{C_p} \frac{1}{{P_0}_{new}} \mbox{ , and } \chi \mbox{ is a number need to be determined.}
$$
Divide Eq. (\ref{eqn:mom}) by $\rho_{new}$ and take divergence
\begin {equation}
\label{eqn:divnew}
\nabla . \vec{u}_{new} = \nabla . \left( \frac{\widetilde{(\rho u_i)}_{new}}{\rho_{new}} \right)-C_{RK4}^{pre}[k] \cdot \Delta t \cdot \nabla . ( \frac{1}{\rho_{new}} \nabla p).
\end{equation}
Now, we can substitute $\nabla . \vec{u}_{new}$ from Eq. (\ref{eqn:divnew}) in Eq. (\ref{eqn:divNewGen})
\begin{equation}
\text{RHS}^{div}_{new}- \chi=\nabla . \left( \frac{\widetilde{(\rho u_i)}_{new}}{\rho_{new}} \right)-C_{RK4}^{pre}[k] \cdot \Delta t \cdot \nabla . ( \frac{1}{\rho_{new}} \nabla p)
\end{equation}
\begin{equation}
\Rightarrow \nabla . ( \frac{1}{\rho_{new}} \nabla p) = \frac{1}{C_{RK4}^{pre}[k] \cdot \Delta t} \left\{ \nabla . \left( \frac{\widetilde{(\rho u_i)}_{new}}{\rho_{new}} \right)-\text{RHS}^{div}_{new}+ \chi\right\} = \text{RHS}_{new}^{Pois} + \chi'.
\end{equation}
This is a variable coefficient Poisson equation. When we take spatial integral of the above equation, for a fully periodic domain, the left hand side term vanishes. Therefore,
\begin{equation}
\chi'=\frac{\chi}{C_{RK4}^{pre}[k] \cdot \Delta t}= \frac{-\int \text{RHS}_{new}^{Pois} dv}{\int dv} \rightarrow \chi = -C_{RK4}^{pre}[k] \cdot \Delta t \cdot \langle{\text{RHS}_{new}^{Pois}}\rangle.
\end{equation}
After solving the variable coefficient Poisson equation, we update the momentum based on Eq. (\ref{eqn:mom}). Note that during the RK4 loop, for a quantity $Q$ with the time evolution equation $\frac{\partial Q}{\partial t}=\text{RHS}_Q$, we construct $Q^{(n+1)}$ as
\begin{equation}
Q^{(n+1)}=Q^{(n+1)} + C_{RK4}^{post}[k] \cdot \Delta t \cdot \text{RHS}_Q [k],
\end{equation}
where $Q^{(n+1)}$ is set to be $Q^{(n)}$ before entering the RK4 loop.
%=====================================================================
%=============================APPENDIX B==============================
%=====================================================================
\section{} \label{sec:appB}
Using a uniform staggered grid, our discretization is kinetic energy conservative. We are solving for momentum components, namely $g_x, g_y$, and $g_z$ (e.g., $g_x = \rho u$). Here, we briefly explain the second-order energy conservative discretization. We use the notation used in Figure \ref{fig:cons} (without losing generality of the 3D case).
For example, consider the following terms for the $x-$momentum equation $\frac{\partial}{\partial x} (\frac{g_x g_x}{\rho}) + \frac{\partial}{\partial y} (\frac{g_x g_y}{\rho}) + \frac{\partial}{\partial z} (\frac{g_x g_z}{\rho})$, where $g_x$, $g_y$, and $g_z$ are momentum in the $x$, $y$, and $z$ directions, respectively. Note that we want all three terms to be computed on the faces (where $x-$momentum variables are stored). 
\begin{equation}
\begin{split}
\Delta x \cdot \frac{\partial}{\partial x} (\frac{g_x g_x}{\rho})_{(I,J)} = \frac{(\frac{{g_x}_{(I+1,J)}+{g_x}_{(I,J)}}{2})(\frac{{g_x}_{(I+1,J)}+{g_x}_{(I,J)}}{2})}{\rho_{(i,j)}} - \\ \frac{(\frac{{g_x}_{(I,J)}+{g_x}_{(I-1,J)}}{2})(\frac{{g_x}_{(I,J)}+{g_x}_{(I-1,J)}}{2})}{\rho_{(i-1,j)}},
\end{split}
\end{equation}
\begin{equation}
\begin{split}
\Delta y \cdot \frac{\partial}{\partial y} (\frac{g_x g_y}{\rho})_{(I,J)} = \frac{({g_x}_{(I,J+1)}+{g_x}_{(I,J)})({g_y}_{(i',j'+1)}+{g_y}_{(i'-1,j'+1)})}{\rho_{(i,j)}+\rho_{(i-1,j)}+\rho_{(i-1,j+1)}+\rho_{(i,j+1)}} - \\ \frac{({g_x}_{(I,J-1)}+{g_x}_{(I,J)})({g_y}_{(i',j')}+{g_y}_{(i'-1,j')})}{\rho_{(i,j)}+\rho_{(i-1,j)}+\rho_{(i-1,j-1)}+\rho_{(i,j-1)}}.
\end{split}
\end{equation}
In 3D, all other terms are computed similarly: first, they are interpolated, and then finite difference is applied.

%=====================================================================
%=============================REFRENCES==============================
%=====================================================================
\bibliographystyle{ctr}

%\bibliographystyle{ctr}
%\bibliography{urzay1.bib}

%%%%%%%%%%%%%%%%%%%%%%%%%%%%%%%%%%%%%%%%%%%%%
\end{document}